\documentclass{elsart}

\usepackage{amssymb}

\begin{document}

\begin{frontmatter}



\title{Electron-ion-ion coincidence experiments for photofragmentation
of polyatomic molecules using pulsed electric fields: treatment of
random coincidences}


\author{G. Pr\"umper, K. Ueda}

\address{Institute of Multidisciplinary Research for Advanced
Materials, Tohoku University, Sendai 980-8577, Japan}

\begin{abstract}
In molecular photofragmentation processes by soft X-rays, a number of ionic fragments can
be produced, each having a different abundance and correlation with the emitted electron
kinetic energy. For investigating these fragmentation processes, electron-ion and
electron-ion-ion coincidence experiments, in which the kinetic energy of electrons are
analyzed using an electrostatic analyzer while the mass of the ions is analyzed using a
pulsed electric field, are very powerful. For such measurements, however, the
contribution of random coincidences is substantial and affects the data in a non-trivial
way. Simple intuitive subtraction methods cannot be applied. In the present paper, we
describe these electron-ion and electron-ion-ion coincidence experiments together with a
subtraction method for the contribution from random coincidences. We provide a
comprehensive set of equations for the data treatment, including equations for the
calculation of error-bars. We demonstrate the method by applying it to the fragmentation
of free CF$_3$SF$_5$ molecules.
\end{abstract}

\end{frontmatter}

\section{Introduction}

This is probably one of the most boring papers ever published in this journal. There may
be no exciting new science in it.  However, it will be one of the most useful papers for
those readers who are doing or planning to do multi-particle coincidence spectroscopy,
because it presents equations in a ``ready-to-use'' form for the treatment of random
coincidences. This problem needs to be solved in order to avoid mistakes in the analysis
on multi-particle coincidence data that are recorded in various kinds of fragmentation
experiments of polyatomic molecules.

Multi-ion coincidence momentum spectroscopy is often used in molecular photofragmentation
studies using synchrotron radiation (SR)~\cite{Lavolee99,Muramatsu02,Ueda05}.
Electron-ion(-ion) coincidence momentum spectroscopy is also widely used in molecular
photoionization study~\cite{Heiser97,Lafosse00,Landers01,Defanis02}. In these
measurements, static extraction fields are applied and all the electrons and ions emitted
in all directions (4$\pi$ sr) are collected. In such measurements, the random
coincidences are minimized by using a sophisticated filtering of the recorded events or
very low ionization rates.

For many applications, however, these approaches and tricks can not be used. Filtering
criteria based on momentum conservation cannot be used if there are some undetected
neutral fragments. This often happens in the reaction pathways for larger molecules. In
such cases, the measured data are no longer kinematically complete. Static extraction
fields used for the above listed experiments spoil the energy resolution for electrons.
Also the measurements are limited to relatively low kinetic energy electrons. If the
kinetic energy of the electron is high (say, more than 50 eV), it becomes practically
impossible to collect the electrons from 4$\pi$ sr keeping reasonable resolution. If one
tries to improve the electron energy resolution, one needs to restrict the acceptance
angle of the electrons, keeping the source volume field-free.  The experiments can be
done, for example, in such a way that the photoelectron emission is observed with limited
acceptance angles in certain directions using an electrostatic analyzer or a conventional
time-of-flight (TOF) analyzer, whereas the molecular fragments emitted all directions
($4\pi$~sr) are collected using pulsed electric
field~\cite{French,Kugeler04,Pruemper05JPB,Rolles05}. In such experiments, the problem of
random coincidences becomes non-trivial, especially if one tries to detect two or more
ions in coincidence with electrons. K\"ammerling and coworkers \cite{Kaemmerling92}
discussed the problem of a high level of electron-random ion coincidences for the case of
multiple photoionization of atoms. They stressed the need for a reference ion measurement
using a ``random'' trigger instead of the electron detection. In a way this paper might
be considered a generalization of their formalism for multiple detectors, if the
multi-hit capability of the ion detector is formally treated as multiple ``yes-no'' ion
detectors. Because of the presence of different fundamental processes for molecules
leading to single ions and multiple ion pairs in the same experiment, the problem is
already rather complicated and we decided to partially neglect the dead time effects of
the ion detector.

In this paper, we describe how to perform these experiments and how to deal with
subtraction of the random coincidences, presenting the method in form of ``ready to
use'' equations. In the following section, we describe an experimental apparatus and
procedures that were successfully used for SR experiments as an practical example.
All the details of the subtraction of the random coincidences will be given in
section 3. In section 4, we present some experimental results as demonstrations for
the necessity of the proper subtraction of the random coincidences. Section 5 is a
summary.

\section{Experiment}

\subsection{Coincidence setup}

Photofragmentation of polyatomic molecules following inner-shell excitation by soft
x-rays has been widely studied in the last two decades. Even for small molecules,
many different ions and ion pairs can be produced, each stemming from different
Auger final states. To disentangle the fragmentation pathways after molecular Auger
decay following inner-shell excitation, coincidence measurement between the Auger
electron and the fragment ion is indispensable. In order to investigate molecular
fragmentation after inner-shell excitation, we developed an
Auger-electron--multiple-ion coincidence apparatus, described in detail
elsewhere~\cite{Pruemper05JPB,Pruemper05JESRP}.  The apparatus was installed at the
c-branch of the high-resolution soft X-ray monochromator~\cite{Ohashi01NIMa} in
beamline 27SU~\cite{Ohashi01NIMb} at SPring-8, an 8 GeV SR facility in Japan: the
radiation source is a figure-8 undulator~\cite{Tanaka96JSR}.

\begin{figure}[!ht]
\vspace{8cm} \includegraphics{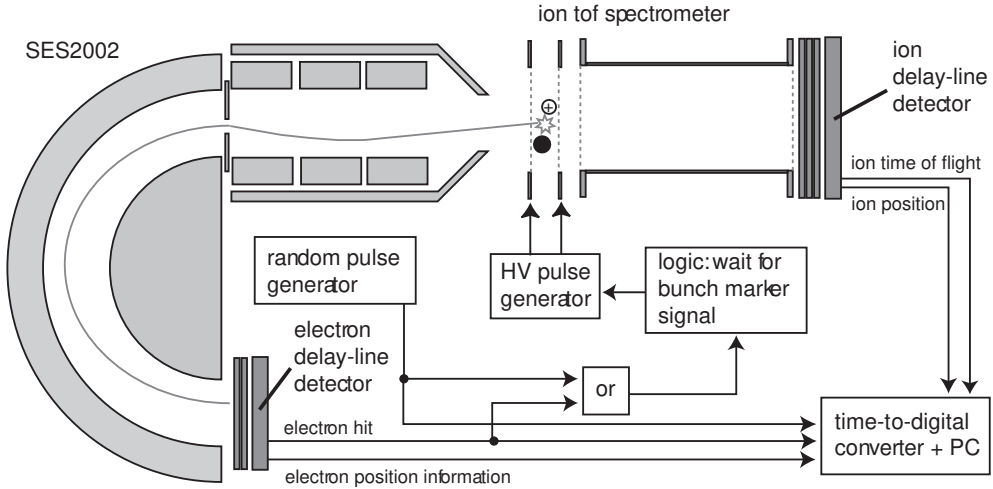} \caption{Schematic of the experimental setup
and the data acquisition system for synchrotron radiation
experiment.} \label{SRsetup}
\end{figure}

A schematic of our apparatus is shown in Fig.~\ref{SRsetup}.  To obtain high
resolution for the electron kinetic energy, a dispersive spectrometer with a limited
detection angle (Gammadata-Scienta SES2002) is used. An ion TOF spectrometer is
placed on the opposite side of the electron spectrometer. The sample gas is
introduced as an effusive beam between the pusher and extractor electrodes of the
ion spectrometer through a needle. No static extraction field is present in the
source region. All voltages of the electron spectrometer are fixed during the
coincidence experiment.  Electrons pass the pusher electrode, enter the electron
spectrometer, and are detected by a fast position resolving detector (Roentdek
DLD40). Triggered by the electron detection or a periodic ``random'' pulse,
rectangular high voltages pulses with opposite signs are generated by a pulse
generator (GPTA HVC-1000) and applied symmetrically to the pusher and extractor
electrodes. The ions are detected by another delay-line detector set (Roentdek
DLD80) at the end of the TOF drift tube. All data are recorded by multichannel
multi-hit time-to-digital converters (Roentdek TDC-8) and stored in the list mode
for off-line analysis. In the list mode, each event consist of one electron or a
random pulse followed by zero, one, two, three or four ions.


\subsection{Optimum conditions for coincidence experiments}

To obtain a reasonable count rate of about 1 Hz a relatively high ionization rate of
some KHz is necessary. Under typical experimental conditions, random coincidences
including one or two ionic fragments that do not come from the same molecule like
the detected electron, are a major contribution to the recorded data. We will only
deal with the case were one electron was detected. Due to the limited acceptance
angle and the limited energy window of the electron spectrometer electron-electron
coincidences are a negligible contribution to the recorded data. The ions that come
from the same molecule like the detected electron will be called ``true ions'' the
others will be called ``random ions''. There is no such distinction for electrons.
There is no way to distinguish between true and random ions for an individual
ionization event. The percentage of the contribution of the random ions must be
determined in an additional measurement. The problem of a huge contribution of
random ions is inherent to the present experiment, as the acceptance angle for the
high resolution electron spectrometer is $\sim 0.001 \cdot$ 4$\pi$ sr and therefore
999 of 1000 electrons are not detected. The corresponding ions need time to escape
from the region seen by the ion spectrometer. A 30 amu ion with a kinetic energy of
0.1~eV takes about 12 $\mu$s to fly 1~cm. So if the double ionization rate via core
hole decay is about 40 KHz, there is at average 1 old ion present, producing a huge
contribution of random coincidences. On the other hand, the electron count rate is
quite low, because the electron spectrometer sees only a limited part of the
spectrum. So in this example, an Auger electron count rate of less than 40 Hz would
occur. As a rule of thumb, the best compromise between count rate and true-to-random
coincides is obtained for 10-30\% random coincidences. In this case, the electron
count rate is typically less than 10 Hz. For ion pairs, the situation can be even
worse. Two ion species can occur with high abundance, but the cross section for the
corresponding ion pair production can be low. In such a case, most of the detected
ions pairs belong to random coincidences. This illustrates the need for a fully
automatic reliable subtraction method for the random coincidence contribution
without any adjustable parameters.

\subsection{Reference measurement for the random coincidences}

A reference measurement for the random coincidences must be performed under exactly
the same experimental conditions as the real coincidence measurement. One very
convenient method is to use not only the electron trigger to extract the ions, but
also a ``random'' trigger that comes from an external pulse generator. In order to
be able to distinguish the two types of triggers, the information about the trigger
type must be recorded as well in the list mode file for each event. Under ideal
experimental conditions the electron trigger will always lead to the detection of an
ion, while the random trigger should never lead to the detection of any ion. Under
typical experimental conditions, however, it is already reasonable, if 1000 electron
triggers lead to the detection of 400 ions, while 1000 random triggers lead only to
the detection of 150 ions. In this case we would assume that about 250 of the ions
that were detected after the electron trigger are true, while 150 of these electrons
are random. So the contribution of the true coincidences is 62.5{\%} in this case.

The random trigger rate can be set higher than the electron count rate. So the
statistical uncertainties of the random contribution are much lower than that of the
real signal. Of course the list mode file should not be overloaded by the random
events. As a rule of thumb, use 10 time as many random triggers, as real electron
triggers.

A synchrotron light source has a repetition rate of several MHz and the time between the
light pulses (time window) is not long enough to detect the electron and the ions.
Therefore, the HV is only triggered when either an electron is detected or a random pulse
is generated by the pulse generator.

Both the true and random ion signals are time-correlated to the synchrotron light.
So is the electron trigger. The random trigger is not. This difference may lead to
subtle differences between the random contribution in the electron triggered events
and the random triggered events. To avoid this difference an electronic trick is
used: There is an electronic signal available, that represents the (delayed) time
structure of the light source. It will be called the bunch marker. During the
measurement the electron and the random trigger signals do not trigger the
extraction field immediately, but wait for the bunch marker signal. The phase of the
bunch marker signal should be adjusted in a way to minimize the additional delay
time after the electron detection. Using this trick, the random events behave
perfectly like the electron events, and the contribution of the random ions can be
subtracted form the total ion signal with the method described in this paper.

\subsection{Experimental methods to minimize the contribution of random coincidences and their limits}

Instead of applying sophisticated statistical methods to subtract the contribution of
random coincidences, it would be much more elegant and straightforward to eliminate them
by a clever experimental design. Two standard techniques exists to minimize the
contribution of the random events. We give a brief account of the limits of their
applicability and show that for the experimental conditions considered here, contribution
of random coincidences cannot be eliminated to a negligible level. The aim of this
article is not to replace these methods, but to properly quantify the contribution of the
remaining random coincidences. Of course a situation where the random contribution can be
neglected never occurs in practice. The reason is very simple: if the contribution of the
random coincidences is reduced by a clever design of the experiment from 10\% to, say
1\%, for a given count rate, this simply means that the light intensity or the gas
pressure can be increased by a factor of 10 to improve the count rate and revert to 10\%
contribution from random coincidences. In other words, an experiment that takes a week
with ``negligible'' random coincidences can be performed in a day if the random
coincidences are subtracted properly.

\begin{figure}[!ht]
\vspace{20cm} \includegraphics{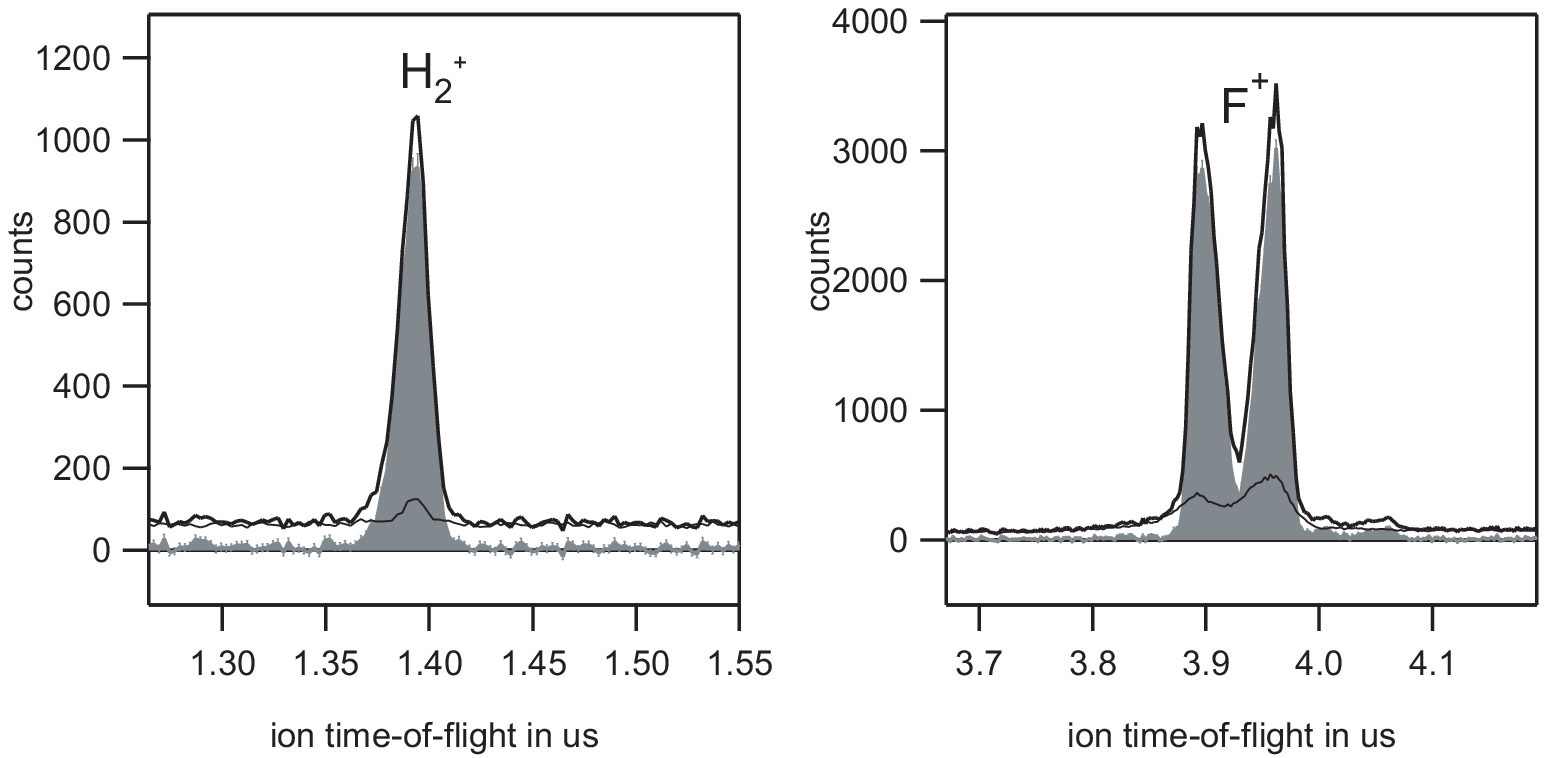}
\caption{Ion time of flight spectrum including slow H$_2$$^+$ and fast F$^+$ ions.
Produced upon F 1s ionization of CH$_3$F. The thick black line is the spectrum produced
from all ions triggered by an electron. The thin line is the contribution of the random
coincidences derived from the spectrum of ions detected after random triggers. The grey
filled graph is the ions spectrum after background subtraction. } \label{I_subtract}
\end{figure}

As described above, slow ions with kinetic energies below 0.1 eV can be the main
contribution to the random coincidences in some experiments. Even if they are created at
a relatively small rate compared to fast ions, the corresponding sharp peaks are clearly
seen in the ion time of flight spectra, because they pile up in the source region until
they are extracted by the high voltage pulse. For example N$_2$$^+$ with a kinetic energy
of 20~meV needs 27$\mu$s to escape from a region with 1 cm radius. When a static
extraction field of 2V/cm is applied, the escape time reduces to about 5 $\mu$s. This
effect is clearly seen in the ion time of flight spectra, the sharp lines from the slow
ions almost disappear. A similar effect can be achieved, by applying a short
($\sim$1$\mu$s) duration HV sweeping pulse every 5$\mu$s. (Of course the electrons
detected during the sweeping pulse should be ignored in the data evaluation.) The obvious
change in the ion spectrum creates the illusion that one can easily control the level
random coincidences by static or pulsed extraction fields. However the contribution of
fast false ions is hardly affected at all. The escape time of F$^+$ with a kinetic energy
of 2 eV is only 2.22 $\mu$s. So the contribution of fake fast F$^+$ ions does not change
significantly using reasonable static extraction fields or sweeping pulses with a
repetition rate lower than 1 MHz. Figure \ref{I_subtract} shows two peaks from the ion
time of flight spectrum of CH$_3$F after F 1s ionization. In this case a small
penetrating field extracted the slow ions, unfortunately in the direction of the ion
detector, leading to flat background in the time of flight spectrum. Nevertheless the
sharp peaks of the slow random ions are efficiently removed in this way. Therefore the
contribution of the random coincidences to the slow H$_2$$^+$ ions is not very large. In
this spectrum the dominating contribution of random ions are fast ions, as can be clearly
seen by the double structure of the peaks belonging to the F$^+$ ions.

Both background suppression methods can only be applied, if the light source used is
either continues or has a very high repetition rate, i.e. discharge lamps or synchrotron
radiation sources, because they are based on the assumption that most random ions are
produced at a different time than the true ions. In experiments with intense pulsed
lasers the repetition rate is typically 1 kHz or less and the experimental conditions are
often such that more than 10\% of the light pulses do ionize at least one molecule. In
such a situation 1\% of the laser pulses may ionize two or more molecules and the
contribution of random coincidences would be about 10\%. In this case the random ions are
produced during the same laser shot in the same small laser focus as the real ions, so
the random ions can not be separated from the real ions using static or pulsed extraction
fields.

\section{Data evaluation method}

\subsection{Preparation}

\subsubsection{The list mode data and filtering}

The data is recorded in the list mode. This means that the data structure is a list of
events, containing an electron position or a random trigger and 0, 1, 2, 3 or 4 ions. The
number 4 is defined by the data acquisition electronics. All ions arriving after the
fourth ion will be ignored. Therefore the formulas in this paper which refer to 4-ion
events, actually mean 4-or-more-ion events. For electron triggered events any of the ions
can come from the same molecule as the electron (true ion) or from another molecule
(false ion). There is no way of telling which ion is true and which is false from the
data. The number of possible combinations of true and false ions are numerous. We will
start the discussion with this event list. Then various numbers and spectra will be
defined by evaluating this event list.

The first step of the data evaluation is filtering. This means that the list mode data is
changed. Filtering can include the selection of the electron energy, the selection of the
ion TOF, the ion emission direction etc. Filtering has two effects. The first effect is
the rejection of unwanted events, such as the events for which the electron energy is not
in the desired interval. The second effect is a modification of events. For example, if
only one ion out of three is in the desired TOF interval, then the event is modified from
a one-electron-three-ion event into a one-electron-one-ion event. To avoid confusion, we
will only refer to the modified event list after this selection process.

\subsubsection{Event statistics}

For a given event list, i.e., for a given choice of the TOF interval for the ions
and energy interval for the electrons, we can do statistics on the events.  We
distinguish events with electron triggers and with random triggers.  The numbers of
these events will be called $N_{e}$ and $N_{RND}$ in the formulas. If more than one
electron was detected (or one electron and a random trigger were present), the event
is not counted.  Since the total number of these events is very small, we do not
discuss the systematic effects of this filtering.

The electron trigger can lead to the detection of 0 to 4 ions. The random trigger can
also lead to to the detection of 0 to 4 ions.  The corresponding probabilities are called
$etP_{0,1,2,3,4}$ and $rtP_{0,1,2,3,4}$. (et stands for electron triggered, rt stand for
random triggered). These numbers are calculated by counting the number of the events and
dividing the result by $N_{e}$ or $N_{RND}$ respectively.

To calculate the random background contribution, we consider the hypothetical
probabilities to detect 0, 1, 2, 3 or 4 ions after one electron detection, when the
background would be absent. These probabilities are called $TP_{0,1,2,3,4}$ ($T$ stands
for true ions). Neglecting the dead time effects of the ion detector, we assume that the
chance to detect a random ion is not effected by the presence of the true ion. So the
chance to detect 0, 1, 2, 3 or 4 false ions after an electron trigger is also
$rtP_{0,1,2,3,4}$.

In order to calculate $TP_{0,1,2,3,4}$ from the known values of $etP_{0,1,2,3,4}$ and
$rtP_{0,1,2,3,4}$, we write down a system of linear equations (\ref{EP_RP_TP}), see the
appendix.  Knowing $rtP_{0,1,2,3,4}$ and $etP_{0,1,2,3,4}$, these equations yield values
for $TP_{0,1,2,3,4}$. The numbers $TP_{0,1,2,3,4}$ defined by these equations should all
be non-negative and they should add up to 1. Because of statistical errors, some of them
may be negative. Therefore the negative values are set to zero and then the numbers
$TP_{0,1,2,3,4}$ normalized to sum up as 1. $TP_0$ is the probability to find no true ion
after the electron detection. There are several possible reasons for this. Most important
one is that the detection efficiency of the ion detector is limited. Second, the choice
of the TOF region for the ions may not contain all ionic fragments. Furthermore, the
electron detector has a certain noise level or dark count rate. So, every
`electron-trigger' does not necessarily correspond to the detection of one electron from
the source region.

\subsubsection{The true number of ionic fragments}

To make statements about the real ionization events without the instrumental influences,
i.e. the detection efficiency, we define five new values, $P_{0,1,2,3,4}$ that represent
the probabilities that 0, 1, 2, 3, 4 ions are present for one electron detected. If the
selected TOF region contains all possible ions, $P_0$ is the probability for an electron
to be detector noise, P$_1$ the probability be an electron leading to one ion, P$_2$ the
probability be an electron from an ionization leading to two ions etc. Please note that
the single and double ionization depend on the charge state of the individual ions. So
$P_1$ and $P_2$ do not exactly belong to single and double ionization processes. To
obtain the probabilities for the single and double ionization, one needs to assign the
charge state to the detected ions. The value of the ion detection efficiency $PD$ is
especially important to compare the yields of doubly charged ions with that of pairs of
singly charged ions, because the latter one suffers form the detection efficiency two
times. Lets assume that the chosen TOF interval contains all possible ions. Further we
assume that the detection efficiency is the same for all ions independent of mass,
momentum and charge state. Its value $PD$ is known (determined form an independent
measurement). Typically $PD$ should be between 0.2 and 0.4 summarizing the transmission
of the grids in the ion spectrometer and the detection efficiency of the detector. So for
given values of PD and $TP_{0,1,2,3,4}$ the values of $P_{0,1,2,3,4}$ can be calculated
from the set of linear equations (\ref{calc_P01234}), given in the appendix.

All values of $P_{0,1,2,3,4}$ must be between 0 and 1. Their sum is 1 by definition. The
value of $PD$ can be estimated in a reference measurement where one of the values $P_1$
or $P_2$ is close to 1. E.g. in the ionization of atoms (e.g. rare gas atoms) ,
neglecting the detector noise $P_1$ is always 1.

If a molecule contains hydrogen, the protons produced are very likely to disobey the
approximation that the detection efficiency is the same for all ionic fragments.
Their detection efficiency is usually smaller than for the other ions, because
protons are usually much faster than the other ions and escape from the region
between the pusher and extractor electrodes, before the high voltage pulse is
applied.

\subsection{Electron spectra}

\subsubsection{All-electron spectrum}

One can make a spectrum of all electrons detected, no matter how many ions where
found in coincidence. This is called all-electron-spectrum $AES(x)$. $x$ stands for
the kinetic energy of the electron, so technically speaking for a detector position.
Because of the discrete nature of the electron detector and the time measurement
using a TDC, $AES(x)$ actually means the chance to detect an electron within an
interval around the integer value $x$. So there is no need to use the notion of
`probability density' for truly continuous spectra. One can create separate electron
spectra $ES_{0,1,2,3,4}(x)$ for the events with 0,1,2,3,4 ions, respectively.
$AES(x)$ is the sum of all these spectra:
\begin{equation}\label{AES}
AES(x)=ES_0(x)+ES_1(x)+ES_2(x)+ES_3(x)+ES_4(x)\;.
\end{equation}
Without the contribution of the random ions, there would be the hypothetical spectra
$HES_{0,1,2,3,4}$(x). The relation between $HES_{0,1,2,3,4}(x)$ and $ES_{0,1,2,3,4}(x)$
is similar to the set of equations~\ref{EP_RP_TP}. One only needs to replace
$etP_{0,1,2,3,4}$ by $ES_{0,1,2,3,4}(x)$ and $TP_{0,1,2,3,4}$ by $HES_{0,1,2,3,4}(x)$.
Therefore $HES_{0,1,2,3,4}(x)$ can be calculated from $ES_{0,1,2,3,4}(x)$. Solving the
equations for $HES_{0,1,2,3,4}(x)$ leads to:
\begin{eqnarray}
HES_0(x) &=& \frac{ES_0(x)}{rtP_0} \nonumber \\
HES_1(x) &=& \frac{ES_1(x)-rtP_1 \cdot HES_0(x)}{rtP_0} \nonumber \\
HES_2(x) &=& \frac{ES_2(x)-rtP_2 \cdot HES_0(x)-rtP_1 \cdot HES_1(x)}{rtP_0} \nonumber \\
HES_3(x) &=& \frac{ES_3(x)-rtP_3 \cdot HES_0(x)-rtP_2 \cdot HES_1(x)-rtP_1 \cdot HES_2(x)}{rtP_0} \nonumber \\
\end{eqnarray}
The corresponding formula for $HES_4$(x) is not written down explicitly, so keep the
formulas more readable.

One should remember that $HES_{0,1,2,3,4}(x)$ are defined as what the spectra would look
like without the random contribution. The existence of the random contribution
redistributes the counts, from the spectra with few ions to the spectra with more ions.
Therefore typically the area of $HES_1(x)$ contains more counts than that of $ES_1(x)$.
In order to have an equation of the form true spectrum = measured spectrum - random
spectrum, we define the true electron spectra in the following way: $TS_0(x)=rtP_0 \cdot
HES_0(x), ..., TES_3(x)=rtP_0 \cdot HES_3(x)$. Then the random backgrounds $BES_1(x)$,
$BES_2(x)$, $BES_3(x)$ can be calculated from $ES_{0,1,2,3}(x)$ in the following way:
\begin{eqnarray}
c_1 &=& \frac{rtP_1}{rtP_0}, c_2 = \frac{rtP_2}{rtP_0}-\frac{rtP_1^2}{rtP_0^2} \nonumber \\
c_3 &=& \frac{rtP_3}{rtP_0}+\frac{rtP_1^2}{rtP_0^2}-2 \cdot \frac{rtP_1 \cdot rtP_2}{rtP_0^2} \nonumber \\
BES_1(x) &=& c_1 \cdot ES_0(x) \nonumber \\
BES_2(x) &=& c_2 \cdot ES_0(x)+c_1 \cdot ES_1(x) \nonumber \\
BES_3(x) &=& c_3 \cdot ES_0(x)+c_2 \cdot ES_1(x)+c_1 \cdot ES_2(x) \nonumber \\
TES_0(x) &=& ES_0(x) \nonumber \\
TES_1(x) &=& ES_1(x)-BES_1(x) \nonumber \\
TES_2(x) &=& ES_2(x)-BES_2(x) \nonumber \\
TES_3(x) &=& ES_3(x)-BES_3(x) \label{TESx_Esx}
\end{eqnarray}

The electron spectrum that belongs to the true e-I coincidences is $TES_1(x)$, the
electron spectrum that belongs to the true e-I-I coincidences is $TES_2(x)$. $TES_3(x)$
belongs to the e-I-I-I coincidences. The different terms that contribute to the
backgrounds $BES_2(x)$ and $BES_3(x)$ have a simple interpretation: $c_2 \cdot ES_0(x)$
is the random background of $TES_2(x)$ that is caused by two false ions, $c_1 \cdot
ES_1(x)$ is the background that is caused by one real and one false ion. $c_3 \cdot
ES_0(x)$ is the background in $TES_3(x)$ that is caused by three false ions $c_2 \cdot
ES_1(x)$ is caused by two false and one real ion, $c_1 \cdot ES2(x)$ is caused by one
false and two real ions. The formulas for the statistical errors of the ture electron
spectra $TES_{0,1,2,3}(x)$ are given by the equations \ref{TES_error} in the appendix.

\subsubsection{Ion specific electron spectra}

For every ion mass to charge ratio, there is a certain small TOF interval called
$Iregion(i)$ that contains these ions. One can do the filtering of the event list by
restricting the accepted TOF interval of the ions to $Iregion(i)$. The electron spectrum
$TES_1(x)$ created after this filtering is called the ion specific electron spectrum
$TES_1I(x,i)$.

\subsection{Ion spectra}

\subsubsection{Ion TOF spectrum and Ion-ion coincidence map}

Similar considerations like for the electron spectra are valid for the ion spectra. The
simplest quantity to consider is the ion TOF spectrum for all ions detected after an
electron trigger $etAI(\hbox{tof})$. (We use ``tof'' instead of ``t'' for the time of
flight to avoid confusion with ``triggered'' and ``true'').

All ions contribute independent if they are detected alone or as part of a (multiple) ion
pair. $tof$ stands for TOF of ion. We will only discuss TOF spectra, but analogous
equations are valid for the mass spectra, created by a non-linear transformation of the
TOF to the mass/charge ratio. A similar spectrum can be generated for all random
triggered ions $rtAI(\hbox{tof})$. Please note that this spectrum does not necessarily
represent the true pattern of fragmentation, because the fast ionic fragments are more
likely to escape from the region between the pusher and extractor electrode than the slow
ones.

Neglecting dead time effects, the detection of one ion is completely independent of the
presence of the other ions. Therefore the background subtraction for $etAI(\hbox{tof})$
is very straightforward. The number of electron triggers and random triggers are
different. Their ratio will be called scaling factor $SC=N_{e}/N_{RND}$.
\begin{eqnarray}
BetAI(\hbox{tof})       &=& SC \cdot rtAI(\hbox{tof})  \nonumber \\
TetAI(\hbox{tof})       &=& etAI(\hbox{tof})-BetAI(\hbox{tof}) \nonumber \\
\Delta TetAI(\hbox{tof}) &=& \sqrt{etAI(\hbox{tof})+SC^2 \cdot rtAI(\hbox{tof})} \nonumber \\
\end{eqnarray}
For the electron-multi ion coincidence technique, the main difference between electron
and ion spectra is, that for single ion events we have a TOF or mass spectrum, for double
ions events we have a two-dimensional (2D) map, etc. We will only consider the cases for
up to two ions. We define $etI(\hbox{tof})$, the spectrum of the ions from single ion
events triggered by electron detection and $etII(\hbox{tof1},\hbox{tof2})$, the 2D ion
TOF spectrum containing the events with two ions triggered by an electron. The ions are
sorted by time of flight, therefore, by definition, we always have the condition
$\hbox{tof1}<\hbox{tof2}$. $rtI(\hbox{tof})$ and $rtII(\hbox{tof1},\hbox{tof2})$ are the
corresponding spectrum and 2D map for the random triggered events, respectively.

The corresponding spectra of true ions are named $TetI(\hbox{tof})$ and
$TetII(\hbox{tof1},\hbox{tof2})$, they are calculated by subtracting the random
coincidences $BetI(\hbox{tof})$ and $BetII(\hbox{tof1},\hbox{tof2})$.

\begin{eqnarray}
TetI(\hbox{tof})       &=& etI(\hbox{tof})-BetI(\hbox{tof}) \nonumber \\
TetII(\hbox{tof1},\hbox{tof2})  &=&
etII(\hbox{tof1},\hbox{tof2})-BetII(\hbox{tof1},\hbox{tof2}) \nonumber \\
BetI(\hbox{tof})       &=& SC \cdot TP_0 \cdot rtI(\hbox{tof}) \nonumber \\
BetII(\hbox{tof1},\hbox{tof2})  &=& SC \cdot TP_0 \cdot rtII(\hbox{tof1},\hbox{tof2}) \nonumber \\
~             &-& \frac{2 \cdot SC \cdot TP_0 \cdot rtI(\hbox{tof1}) \cdot rtI(\hbox{tof2})}{rtP_0 \cdot N_{RND}}   \nonumber \\
~             &+& \frac{etI(\hbox{tof1}) \cdot rtI(\hbox{tof2}) + rtI(\hbox{tof1}) \cdot
etI(\hbox{tof2})}{rtP_0 \cdot N_{RND}}
\end{eqnarray}

The explanation for these formulas is given in in the appendix, see formulas
\ref{Etit1_PEti_t1} and \ref{TetII_def}.

\subsubsection{Ion pair statistics}

For each ion pair $IIpair(j)$ there is a specific region $IIregion(j)$ in the
$\hbox{tof1}-\hbox{tof2}$ plane. Because the possible momentum correlation, e.g.,
emission into opposite directions in a two body fragmentation, these regions are not
necessarily chosen rectangular, but rather like tilted ellipses. The number of counts for
an ion pair is defined by the ion-ion maps:
\begin{eqnarray}
TCtsIIpair(j) &=& \sum^{\hbox{tof1},\hbox{tof2}}_{in~IIregion(j)} TetII(\hbox{tof1},\hbox{tof2}) \nonumber \\
CtsIIpair(j)  &=& \sum^{\hbox{tof1},\hbox{tof2}}_{in~IIregion(j)} etII(\hbox{tof1},\hbox{tof2}) \nonumber \\
BCtsIIpair(j) &=& \sum^{\hbox{tof1},\hbox{tof2}}_{in~IIregion(j)} BetII(\hbox{tof1},\hbox{tof2}) \nonumber \\
TCtsIIpair(j) &=& CtsIIpair(j)-BCtsIIpair(j)
\end{eqnarray}
The last equation is useful to assign error-bars to the number of
true ion pairs
\begin{equation}\label{IIpairdelta}
\Delta TCtsIIpair(j) = \sqrt{CtsIIpair(j)+BCtsIIpair(j)}
\end{equation}
For the sake of simplicity, we did not use a proper error propagation formula for
$BCtsIIpair(j)$. Instead, we simply considered $\sqrt{BCtsIIpair(j)}$ a good estimate of
the statistical uncertainties of $BCtsIIpair(j)$. Strictly speaking, this is an
underestimate of the statistical error, if the random background in the single ion
spectra $etI(\hbox{tof})$ is large, due to poor experimental conditions. In that case
\begin{equation}\label{IIpairdelta2}
\Delta TCtsIIpair(j) = \sqrt{CtsIIpair(j)+\sqrt{2} \cdot BCtsIIpair(j)}
\end{equation}
can be used as an upper limit for $\Delta TCtsIIpair(j)$.

\subsection{e-I coincidence diagrams}

From the events containing exactly one electron and one ion, three diagrams are created.
Two of them have been defined before. An electron spectrum $ES_1(x)$ or the true electron
spectrum for single ions: $TES_1(x)$. The ion spectrum for the single ion events,
$etI(\hbox{tof})$ or the true single ion events $TetI(\hbox{tof})$. Now we define a 2D
map $etEI(x,\hbox{tof})$ from the electron-ion-events. The total number of the events
coincides with the number of counts in $ES_1(x)$ and in $etI(\hbox{tof})$. One can check
the consistence of the definitions so far, and find that also the number of counts in the
true electron spectrum $TES_1(x)$ and the true ion spectrum $TetI(\hbox{tof})$ is the
same. Now we define the hypothetical quantity $HetEI(x,\hbox{tof})$, the 2D map from the
EI events, if there where no random ions. In our formalism ions can be real or false,
depending if they belong to the same ion as the electron. According to this definition,
the words `real' and `false' cannot be applied to an electron: there are no false
electrons. (Detector noise is considered an electron count without real ions.) So,
$etEI(x,\hbox{tof})$ contains only two contributions: pairs of electrons and real ions
and pairs of electrons and false ions. $HetEI(x,\hbox{tof})$ is related to the measured
map $etEI(x,\hbox{tof})$ by:
\begin{eqnarray}
\frac{etEI(x,\hbox{tof})}{N_{e}}     &=& rtP_0 \cdot \frac{HetEI(x,\hbox{tof})}{N_{e}} + \frac{HES_0(x)}{N_{e}} \cdot \frac{rtI(\hbox{tof})}{N_{RND}} \nonumber \\
\end{eqnarray}
$etEI(x,\hbox{tof})/N_{e}$ is the chance to detect an electron with the energy x and and
ion with the TOF t. $HetEI(x,\hbox{tof})/N_{e}$ is the chance to detect a real ion with
the TOF $tof$ and an electron with the energy $x$. $rtP_0$ is the chance to detect no
further ion. The second term describes the random contribution. $HES_0(x)/N_{e}$ is the
chance to detect an electron with the energy $x$ without a real ion.
$rtI(\hbox{tof})/N_{RND}$ is the chance to detect a single false ion with the TOF $tof$.

We substitute $HES_0(x)=TES_0(x)/rtP_0=ES_0(x)/rtP_0$ and define the true electron-ion
map as $TetEI(x,\hbox{tof})=rtP_0 \cdot HetEI(x,\hbox{tof})$
\begin{eqnarray}
etEI(x,\hbox{tof})     &=& TetEI(x,\hbox{tof}) + \frac{ES_0(x) \cdot rtI(\hbox{tof})}{rtP_0 \cdot N_{RND}} \nonumber \\
\end{eqnarray}
and get the random contribution to etEI(x,\hbox{tof})
\begin{eqnarray}
BetEI(x,\hbox{tof})    &=& \frac{ES_0(x) \cdot rtI(\hbox{tof})}{rtP_0 \cdot N_{RND}}   \nonumber \\
TetEI(x,\hbox{tof})     &=& etEI(x,\hbox{tof}) - BetEI(x,\hbox{tof}) \nonumber \\
\end{eqnarray}
The area of $BetEI(x,\hbox{tof})$ coincides with the area of $BES_1(x)$ and
$BetI(\hbox{tof})$. It is always the number of false e-I coincidences.

\subsubsection{A remark about ion specific electron spectra}

$TetEI(x,\hbox{tof})$ contains all information about the electrons detected in
coincidence with specific ions. So the corresponding ion specific electron spectra
$TES_1I(x,i)$ could be defined as $\sum^{tof}_{in~Iregion(i)} TetEI(x,\hbox{tof})$. This
method has a big disadvantage over the method based on filtering the event list,
described above. Only the events with exactly one ion contribute to $etEI(x,\hbox{tof})$
-- and therefore to $TetEI(x,\hbox{tof})$. So, usually the statistics in $TES_1I(x,i)$
get better when $TES_1I(x,i)$ is defined as $TES_1(x)$ when the time-flight-range of the
ions is restricted to $Iregion(i)$ in the filtering of the event file. So the multiple
ion events with one ion in the interval $Iregion(i)$ become single ion events and
contribute to $TES_1I(x,i)$.

\subsubsection{Ion-pair specific electron spectra}

Now we will discuss how to determine the random background for the spectra of electrons
that are coincident with a specific pair of ions. Therefore we formally introduce the 3D
map $etEII(x,\hbox{tof1},\hbox{tof2})$ that describes the distribution of the
electron--ion-pair events. It is related to $etII(\hbox{tof1},\hbox{tof2})$ and to
$ES_2(x)$ by:
\begin{eqnarray}
etII(\hbox{tof1},\hbox{tof2})    &=& \sum_{all~x} etEII(x,\hbox{tof1},\hbox{tof2})   \nonumber \\
ES_2(x)         &=& \sum_{all~\hbox{tof1},\hbox{tof2}} etEII(x,\hbox{tof1}\hbox{tof2})
\end{eqnarray}
There will be no need to actually create this 3D array $etEII(x,\hbox{tof1},\hbox{tof2})$
in the computer's memory. The results we will present, only make use of 2D arrays.

The electron spectrum that belongs to a specific ion pair is called
$ES_2IIpair(x,j)$. It is the histogram of all electrons coincident
with two ions in the region $IIregion(j)$.
\begin{eqnarray}
ES_2IIpair(x,j)  &=& \sum^{\hbox{tof1},\hbox{tof2}}_{in~IIregion(j)}
etEII(x,\hbox{tof1},\hbox{tof2})
\end{eqnarray}
Lets define the hypothetical 3D spectrum $HetEII(x,\hbox{tof1},\hbox{tof2})$ that would
be there without the contribution of the random coincidences. Similar to the
considerations for $etII(\hbox{tof1},\hbox{tof2})$ the relations between
$HetEII(x,\hbox{tof1},\hbox{tof2})$ and $etEII(x,\hbox{tof1},\hbox{tof2})$ these spectra
is:
\begin{eqnarray}
\frac{etEII(x,\hbox{tof1},\hbox{tof2})}{N_{e}}  &=& rtP_0 \cdot \frac{HetEII(x,\hbox{tof1},\hbox{tof2})}{N_{e}} \nonumber \\
                &+& \frac{HetEI(x,\hbox{tof1})}{N_{e}} \cdot \frac{rtI(\hbox{tof2})}{N_{RND}} + \frac{rtI(\hbox{tof1})}{N_{RND}} \cdot \frac{HetEI(x,\hbox{tof2})}{N_{e}} \nonumber \\
                &+& \frac{rtII(\hbox{tof1},\hbox{tof2})}{N_{RND}} \cdot  \frac{HES_0(x)}{N_{e}}
\end{eqnarray}
Meaning that the electron that is detected in coincidence with two ions, can be detected
with two true ions, one true and one false ion, one false and one true ion or a pair of
false ions. We define the corresponding true electron-ion-ion map
$TetEII(x,\hbox{tof1},\hbox{tof2})=rtP_0 \cdot HetEII(x,\hbox{tof1},\hbox{tof2})$ and
get:
\begin{eqnarray}
TetEII(x,\hbox{tof1},\hbox{tof2}) &=& etEII(x,\hbox{tof1},\hbox{tof2}) \nonumber \\
                &-& \frac{TetEI(x,\hbox{tof1}) \cdot rtI(\hbox{tof2})}{rtP_0 \cdot N_{RND}} - \frac{rtI(\hbox{tof1}) \cdot TetEI(x,\hbox{tof2})}{rtP_0 \cdot N_{RND}} \nonumber \\
                &-& rtII(\hbox{tof1},\hbox{tof2}) \cdot \frac{SC}{rtP_0} \cdot \frac{ES_0(x)}{N_{e}}
\end{eqnarray}
The true electron spectrum coincident with true ions in the region
$IIregion(j)$ is defined by:
\begin{eqnarray}
TES_2IIpair(x,j)  &=& \sum^{\hbox{tof1},\hbox{tof2}}_{in~IIregion(j)} TtEII(x,\hbox{tof1},\hbox{tof2})   \nonumber \\
\end{eqnarray}
So we define the corresponding random background $BS_2IIpair(x,j)$:

\begin{eqnarray}
BES_2IIpair(x,j)  &=& \sum^{\hbox{tof1},\hbox{tof2}}_{in~IIregion(j)} \frac{TetEI(x,\hbox{tof1}) \cdot rtI(\hbox{tof2})+rtI(\hbox{tof1}) \cdot TetEI(x,\hbox{tof2})}{rtP_0 \cdot N_{RND}}   \nonumber \\
                &+& \frac{ES_0(x)}{N_{e}} \sum^{\hbox{tof1},\hbox{tof2}}_{in~IIregion(j)}  rtII(\hbox{tof1},\hbox{tof2}) \cdot \frac{SC}{rtP_0}    \nonumber \\
TES_2IIpair(x,j)  &=& ES_2IIpair(x,j) - BES_2IIpair(x,j) \nonumber \\
\Delta TS_2IIpair(x,j) &=& \sqrt{ES_2IIpair(x,j)+BES_2IIpair(x,j)}
\end{eqnarray}

If the summation region $IIregion(j)$ is extended to all ion-ion pairs
($\hbox{tof1}<\hbox{tof2}$) $BES_2IIpair(x,j)$ becomes equal to $BES_2(x)$.

There is no way to simplify the equation for $BES_2IIpair(x,j)$ any further. The
summation over all combinations of $\hbox{tof1}$ and $\hbox{tof2}$ in the II-region must
actually be carried out. This shows that in general the correlation of the true ion and
electron (described by $TetEI(x,\hbox{tof})$) and the correlation of two false ions
(described by $rtII(\hbox{tof1},\hbox{tof2})$) must be considered for the background
subtraction. For the sake of simplicity, in the calculation of $\Delta TS_2 IIpair(x,j)$
we did not use a proper error propagation formula for $BES_2IIpair(x,j)$. We simply
considered $\sqrt{BES_2IIpair(x,j)}$ a good estimate of the statistical uncertainties of
$BES_2IIpair(x,j)$. For reasonable experimental conditions, i.e. if the area of
TetEI(x,\hbox{tof}) is larger than the area of BetEI(x,\hbox{tof}) in the relevant
regions, this is a good estimate.

\subsection{Angular distributions, KER spectra}

In this paper, we only consider energy spectra of electrons and mass
spectra of ions and their correlations. Exactly the same formalism
can be applied to distributions with respect to any variable, e.g.,
emission angles or molecular frame fixed angles, linear momentum,
kinetic energy release etc. All of this quantities have to be
calculated event by event from the detection positions and the
detection times of the particles.

\section{Experimental results}

\begin{figure}[!ht]
\vspace{20cm} \includegraphics{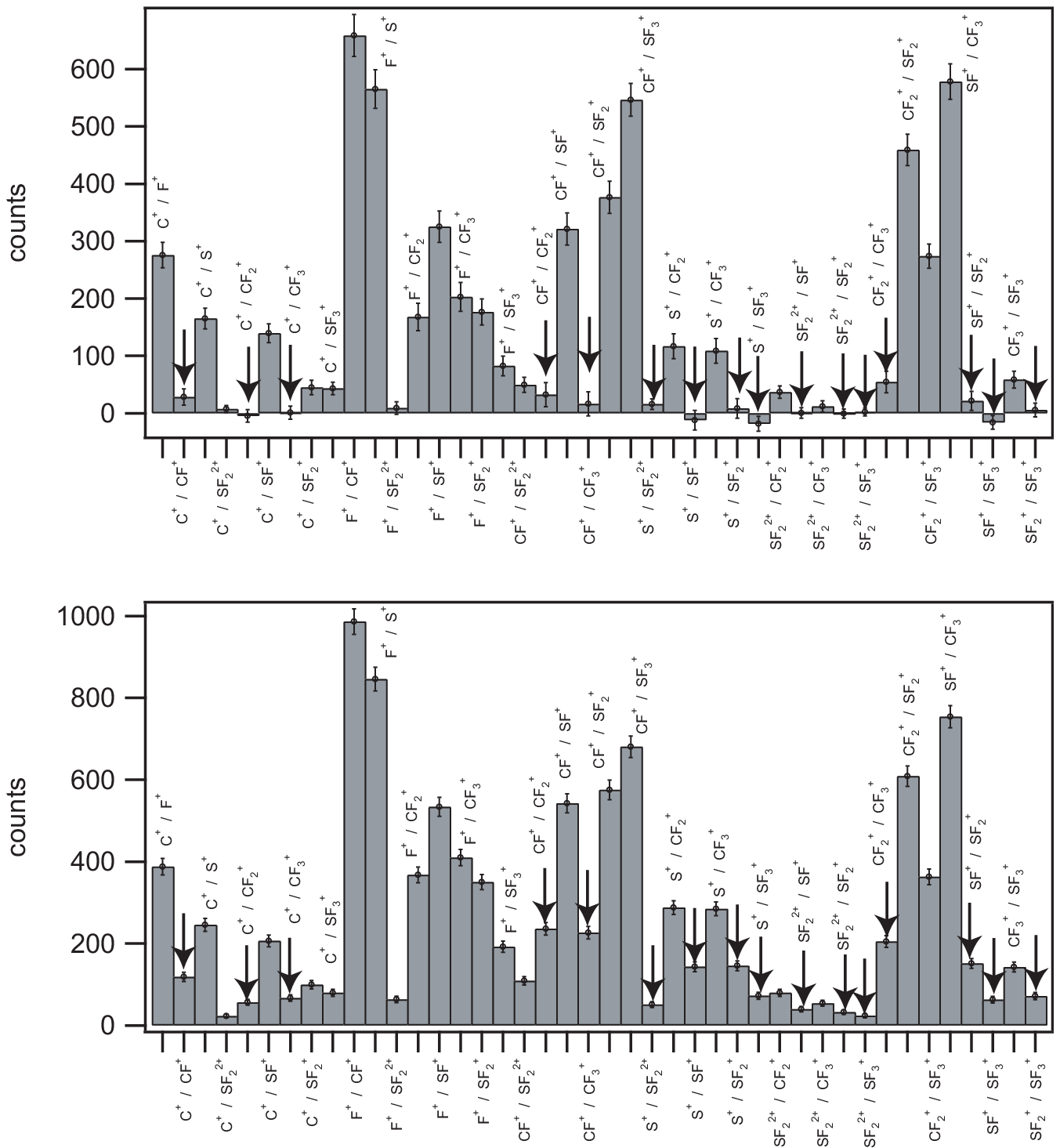}
\caption{Result of the subtraction for ion pairs. No subtraction lower panel, with
subtraction upper panel. The arrows indicate impossible ion combinations.}
\label{II_subtract}
\end{figure}

As a test for the procedures we analyzed the coincidence data of the molecule
SF$_5$CF$_3$ taken at a photon energy of 746.95~eV. The chemically shifted F(1s)
photo lines where recorded in coincidence with ion pairs. This data set is an ideal
test for the methods described here, because F(1s) ionization leads almost
inevitably to the production of at least two ions. The molecule does not contain any
hydrogen, which usually suffers from low detection efficiency in this type of
experiment. Most importantly, the molecule contains only one carbon and one sulfur
atom. Therefore a lot of forbidden ion fragment combinations, containing either two
carbon or two sulfur atoms appear in the data before the subtraction of the random
ion contributions. We know that the contribution from these combinations must be
exactly zero after the subtraction of the random coincidences.
Figure~\ref{II_subtract} shows the abundance of the most prominent ion pairs. Among
the 42 ion pairs, 16 pairs are forbidden, indicated by the arrows. Without the
subtraction  of the random contribution (lower panel), some of them e.g.
CF$^+$-CF$_2$$^+$, CF$^+$-CF$_3$$^+$, CF$_2$$^+$-CF$_3$$^+$, and SF$^+$-SF$_2$$^+$
have intensities comparable to the dominating allowed peaks. After the subtraction
10 of the 16 forbidden pairs (62 percent) have zero intensity, within a 1 sigma
error-bar, and except SF$_2$$^+$-SF$_3$$^+$ (2.5 sigma) all of then are zero within
a two sigma error-bars. The error-bars were calculated using equation
\ref{IIpairdelta}. After the random subtraction no negative intensities occur. So
there is also no overestimation of the random background. You can also see, that
some of the allowed ions pairs get zero intensity after the subtraction, e.g,
SF$_2$$^{2+}$-CF$_3$$^+$. Without the subtraction of the random background these ion
pairs would have been considered real.

\begin{figure}[!ht]
\vspace{20cm} \includegraphics{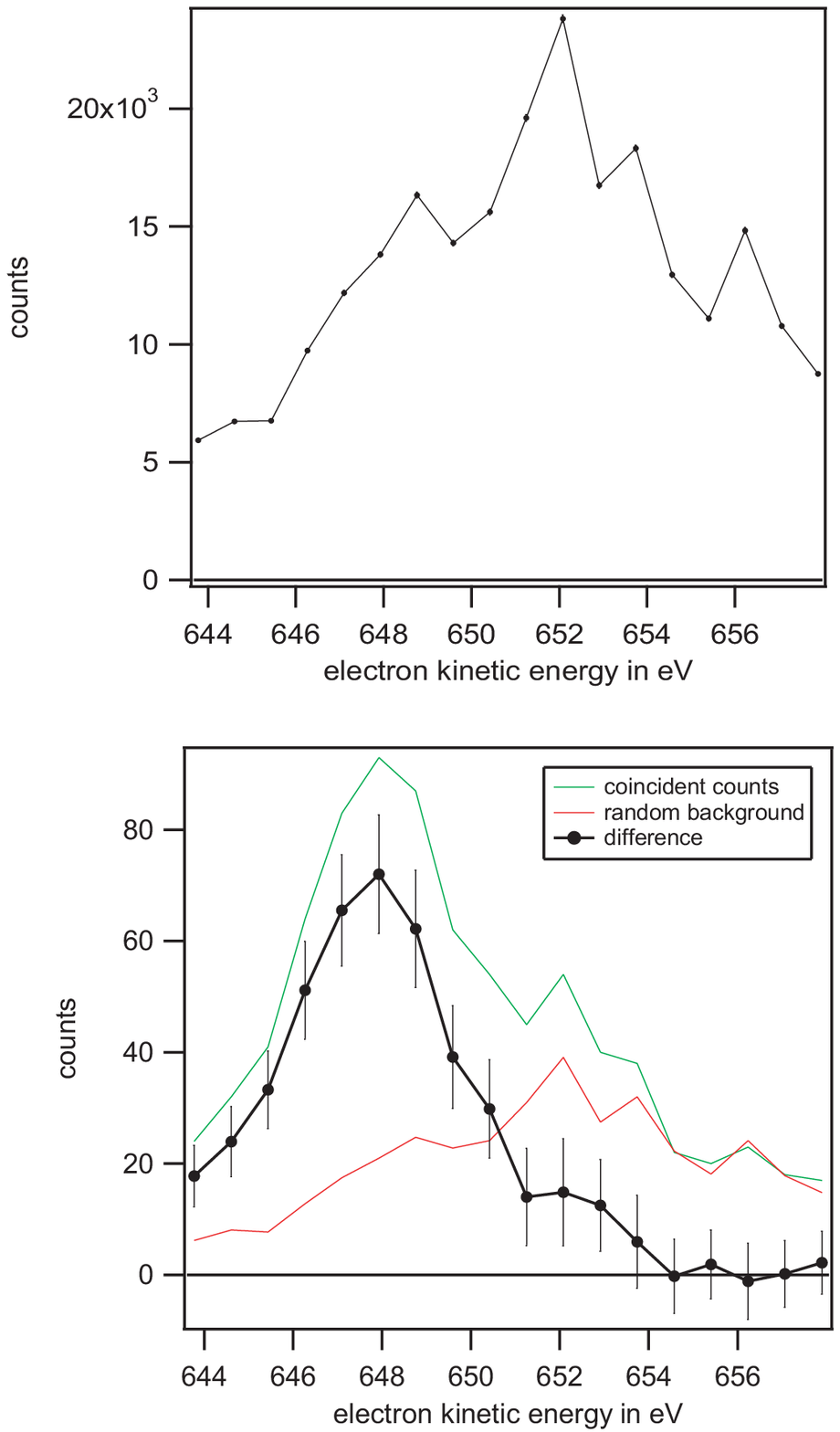} \caption{Result of the subtraction for ion pairs in the resonant Auger
spectrum at a photon energy of 787.9~eV. Here we have a very low level of ion pair
production. Upper panel, total electron spectrum, Lower panel, electrons coincident
with ion pairs.} \label{II_subtract2}
\end{figure}

The second test was even more strict. We set the photon energy to 687.9~eV and
recorded the resonant Auger electrons in the kinetic energy window from 644~eV to
657~eV. The contribution from ion pairs was very low. Only two ions pairs were found
to contribute CF$_2$$^+$ - SF$_3$$^+$ and CF$_3$$^+$ - SF$_3$$^+$. Even though the
ion pair production amounts in less than 1 percent of the events, the shape of
electron spectrum coincident with these ion pairs, is clearly different from the
total electron spectrum. After the subtraction of the random background it shows
zero intensity for kinetic energies above 653~eV, in sharp contrast to the
non-coincident spectrum. These two examples clearly illustrate the non-trivial
effects due to the contribution of the random coincidences that can easily lead to
severe misinterpretation of the data, if not considered correctly.

\section{Conclusion}

A comprehensive method for the treatment of random coincidence events that appear in
electron-ion-ion coincidence experiments is given and tested for the fragmentation
of SF$_5$CF$_3$ following F(1s) core excitation. The method produces reliable
results even for high levels of random coincidences. The method does not contain any
adjustable parameters and removes all forbidden ion pairs completely. This method is
now used as a standard tool in the data evaluation of PEPIPI-coincidence data sets
taken at the gas phase beamline SU27 at SPring8 in Japan. These data sets include
large molecules with hundreds of possible ion pair combinations. The same method can
be used for laser based multi-photon ionization experiments. For high peak powers
$>10^14$ watt/cm$^2$ multiple ionization becomes the main ionization process for
molecules. Under typical experimental conditions, even for low target density there
is more than one molecule ionized per laser shot, leading to a substantial
contribution of random coincidences.

\section{Acknowledgements}

The measurements were carried out with the approval of the SPring-8 program review
committee and were supported in part by Grants-in-Aid for Scientific Research from
the Japan Society for the Promotion of Science. The authors are grateful to the
staff of SPring-8 for their help in the course of these studies.

\section{Appendix}

\subsection{How to calculate $TP_{0,1,2,3,4}$}

The chance to find no ion after an electron trigger, $etP_0$, is the chance to find no
true ion $TP_0$ and no random ion $rtP_0$; the chance to find exactly one ion is the
chance to find a true ion and no random ion plus the chance to find no true ion and
exactly one random ion; and so on.
\begin{eqnarray}
etP_0 &=& rtP_0 \cdot TP_0 \nonumber \\
etP_1 &=& rtP_1 \cdot TP_0+rtP_0 \cdot TP_1 \nonumber \\
etP_2 &=& rtP_2 \cdot TP_0+rtP_1 \cdot TP_1+rtP_0 \cdot TP_2 \nonumber \\
etP_3 &=& rtP_3 \cdot TP_0+rtP_2 \cdot TP_1+rtP_1 \cdot TP_2+rtP_0 \cdot TP_3 \nonumber \\
etP_4 &=& \sum^{i,j}_{i+j\geq 4} rtP_i \cdot TP_j \label{EP_RP_TP}
\end{eqnarray}
\subsection{How to calculate $P_{0,1,2,3,4}$}

 The probability not to detect an ion is $\overline{PD} = 1 - PD$. For example, the
chance to detect exactly 3 true ions $TP_3$ is the chance that exactly 3 true ions are
present $P_3$ times the chance to detect all of them $PD^3$ plus the chance that 4 true
ions are present times the chance to detect exactly 3 out of 4 ions $4 \cdot
\overline{PD} \cdot PD^3$. So for given values of PD and $TP_{0,1,2,3,4}$ the values of
$P_{0,1,2,3,4}$ can be calculated from the following set of linear equations:

\begin{eqnarray}
TP_0 &=& P_0 + P_1 \cdot \overline{PD} + P_2 \cdot \overline{PD}^2 + P_3 \cdot \overline{PD}^3 + P_4 \cdot \overline{PD}^4 \nonumber \\
TP_1 &=& P_1 \cdot PD + 2 \cdot P_2 \cdot PD \cdot \overline{PD} + 3 \cdot P_3 \cdot \overline{PD}^2 \cdot PD + 4 \cdot P_4 \cdot \overline{PD}^3 \cdot PD \nonumber \\
TP_2 &=& P_2 \cdot PD^2 + 3 \cdot P_3 \cdot \overline{PD} \cdot PD^2 + 6 \cdot P_4 \cdot \overline{PD}^2 \cdot PD^2 \nonumber \\
TP_3 &=& P_3 \cdot PD^3 + 4 \cdot P_4 \cdot \overline{PD} \cdot PD^3 \nonumber \\
TP_4 &=& P_4 \cdot PD^4 \label{calc_P01234}
\end{eqnarray}

\subsection{How to calculate the statistical errors of $TES_{0,1,2,3}(x)$}

$rtP_{0,1,2,3,4}$ are known with good statistics. Their error and the error of
$c_{1,2,3}$ can be neglected, so the statistical errors of the electron spectra that
belong to the true coincidences can be calculated using the following equations:
\begin{eqnarray}
\Delta TES_0(x) &=& \sqrt{ES_0(x)} \nonumber \\
\Delta TES_1(x) &=& \sqrt{ES_1(x)+c_1^2 \cdot ES_0(x)} \nonumber \\
\Delta TES_2(x) &=& \sqrt{ES_2(x)+c_2^2 \cdot ES_0(x)+c_1^2 \cdot ES_1(x)} \nonumber \\
\Delta TES_3(x) &=& \sqrt{ES_3(x)+c_3^2 \cdot ES_0(x)+c_2^2 \cdot ES_1(x)+c_1^2 \cdot
ES_2(x)} \label{TES_error}
\end{eqnarray}
\subsection{How to calculate $TetI(\hbox{tof})$ and $TetII(\hbox{tof1},\hbox{tof2})$}

 Let us define the hypothetical spectra $HetI(\hbox{tof})$ and
$HetII(\hbox{tof1},\hbox{tof2})$ that would be there without the contribution of the
random coincidences. If an ion is detected after an electron trigger, it can be a real or
a false ion, i.e. it can come from the same molecule like the electron or not. If two
ions A and B are detected, there are four possible cases, 1) A is real and B is real, 2)
A is real and B is false, 3) A is false and B is real, 4) A is false and B is false. So
we have the following relation between the ion spectra.
\begin{eqnarray}
\frac{etI(\hbox{tof1})}{N_{e}}     &=& rtP_0 \cdot \frac{HetI(\hbox{tof1})}{N_{e}} + TP_0 \cdot \frac{rtI(\hbox{tof1})}{N_{RND}} \nonumber \\
\frac{etII(\hbox{tof1},\hbox{tof2})}{N_{e}} &=& rtP_0 \cdot \frac{HetII(\hbox{tof1},\hbox{tof2})}{N_{e}} \nonumber \\
~                &+& \frac{HetI(\hbox{tof1})}{N_{e}} \cdot \frac{rtI(\hbox{tof2})}{N_{RND}} + \frac{rtI(\hbox{tof1})}{N_{RND}} \cdot \frac{HetI(\hbox{tof2})}{N_{e}} \nonumber \\
~                &+& TP_0 \cdot
\frac{rtII(\hbox{tof1},\hbox{tof2})}{N_{RND}}\label{Etit1_PEti_t1}
\end{eqnarray}
The first equation is easy to understand. $etI(\hbox{tof1})/N_{e}$ is the chance to find
exactly one ion and this ion has a TOF close to $\hbox{tof1}$ after an electron trigger.
$rtP_0 \cdot HetI(\hbox{tof1})/N_{e}$ is the chance to find no false ion and a true ion
with the TOF close to $\hbox{tof1}$. $TP_0 \cdot rtI(\hbox{tof1})/N_{RND}$ is the chance
to find no true ion and a false single ion with the TOF close to $\hbox{tof1}$.

Now let us consider the second equation. $etII(\hbox{tof1},\hbox{tof2})/N_{e}$ is the
chance to find exactly two ions and the first ion has a TOF close to $\hbox{tof1}$ and
the second ion has a TOF close to $\hbox{tof2}$. $RP_0 \cdot
HetII(\hbox{tof1},\hbox{tof2})/N_{e}$ is the chance to detect no false ion and one real
ion with time of flight $\hbox{tof1}$ and a second real ion with time of flight
$\hbox{tof2}$. $HetI(\hbox{tof1})/N_{e}$ is the chance to detect a real ion with the time
of flight $\hbox{tof1}$ and no further real ion, $rtI(\hbox{tof2})/N_{RND}$ is the chance
to detect a random ion with the TOF $\hbox{tof2}$ and no further random ion, the third
term is the same, with the two ions exchanged, $TP_0 \cdot
rtII(\hbox{tof1},\hbox{tof2})/N_{RND}$ is the chance detect no true ion and to detect a
pair of random ions with the times, $\hbox{tof1}$ and $\hbox{tof2}$ and no further random
ion. Analogue to the case of the true electron spectra $TES_{0,1,2,3,4}(x)$ we define
$TetI(\hbox{tof})=rtP_0 \cdot HetI(\hbox{tof}), TetII(\hbox{tof1},\hbox{tof2})=rtP_0
\cdot HetII(\hbox{tof1},\hbox{tof2})$. We call $TetI(\hbox{tof})$ and
$TetII(\hbox{tof1},\hbox{tof2})$ the true ion intensities. Using these definitions, the
equations~\ref{Etit1_PEti_t1} can be solved for $TetI(\hbox{tof})$ and
$TetII(\hbox{tof1},\hbox{tof2})$:
\begin{eqnarray}
BetI(\hbox{tof})       &=& SC \cdot TP_0 \cdot rtI(\hbox{tof}) \nonumber \\
TetI(\hbox{tof})       &=& etI(\hbox{tof})-BetI(\hbox{tof}) \nonumber \\
BetII(\hbox{tof1},\hbox{tof2})  &=& SC \cdot TP_0 \cdot rtII(\hbox{tof1},\hbox{tof2}) \nonumber \\
~             &-& \frac{2 \cdot SC \cdot TP_0 \cdot rtI(\hbox{tof1}) \cdot rtI(\hbox{tof2})}{rtP_0 \cdot N_{RND}}   \nonumber \\
~             &+& \frac{etI(\hbox{tof1}) \cdot rtI(\hbox{tof2}) + rtI(\hbox{tof1}) \cdot etI(\hbox{tof2})}{rtP_0 \cdot N_{RND}} \nonumber \\
TetII(\hbox{tof1},\hbox{tof2})  &=&
etII(\hbox{tof1},\hbox{tof2})-BetII(\hbox{tof1},\hbox{tof2}) \label{TetII_def}
\end{eqnarray}
To assign an error for $TetI(\hbox{tof})$ , we neglect the error in $SC$ and $TP_0$.
\begin{eqnarray}
\Delta TetI(\hbox{tof}) &=& \sqrt{etI(\hbox{tof})+SC^2 \cdot TP_0^2 \cdot rtI(\hbox{tof})} \nonumber \\
\end{eqnarray}

\subsection{Detector efficiency correction for the electrons and dead
time effects for the ions}

The detection efficiency of the electron detector depends on the detector position, i.e.
on the energy of the electron. Therefore a correction is necessary. We define a
correction function $e\_corr(x)$ whose values vary around 1. It is defined as the average
detection efficiency divided by the position dependent detection efficiency. It can be
applied in the very end to all electron spectra and the electron-ion coincidence map. It
does not affect the data treatment mentioned before. So, for example, $TES_2IIpair(x,j)$
is simply multiplied by $e\_corr(x)$ point by point.

The value of the dead time of the ion detector $DT$ and the subsequent electronics can
easily be determined from an analysis of the time difference of all recorded ion pairs.
In practice, the dead time effects of the ion detector have the strongest effect on the
ion spectrum containing all ions, $TetAI(\hbox{tof})$. They can lead to an overestimation
of $BetAI(\hbox{tof})$ and thus to negative intensity behind a very intense peak in
$TetAI(\hbox{tof})$. For a given detection time $tdet$ and a known dead time $DT$, the
chance for a dead detector $PDT(tdet)$ due to the true ions is the number of electron
triggered true ions inside the detection time interval [$tdet-DT,tdet$] divided by the
number of electron triggers. The probability to find the detector alive is given by:
$Palive(tdet)=1-PDT(tdet)$. So the formula for the background subtraction including the
dead-time is:
\begin{eqnarray}
BetAI(\hbox{tof})       &=& SC \cdot rtAI(\hbox{tof}) \cdot Palive(\hbox{tof}) \nonumber \\
TetAI(\hbox{tof})       &=& etAI(\hbox{tof})-BetAI(\hbox{tof}) \nonumber \\
\end{eqnarray}
As the calculation of $Palive(\hbox{tof})$ requires the knowledge of $TetAI(\hbox{tof})$
and vise versa, one has to use an iterative method to obtain both. So in the first loop,
set $Palive(\hbox{tof})$ equal to 1. Usually three iteration loops are sufficient.

The dead time has not only an effect for the background subtraction of
$TetAI(\hbox{tof})$ but on most other spectra as well. Normally this effect can be
neglected. For the sake of completeness we give a brief recipe how to treat the dead time
effects in a simple way. We assumed that the chance to detect a random ion is not
effected by the presence of the true ions. For high count rates, this is not the case and
the values of $rtP_{0,1,2,3,4}$ redistribute because of dead time effects, so that the
lower ion numbers become more dominant.

Now we go back to the approximation made earlier. We assumed that the detection
probability of a random ion after an electron trigger is equal to the detection
probability of a random ion after a random trigger. Knowing $Palive(\hbox{tof})$ we can
have a better estimate of this: So far we assumed that the
\begin{equation}\label{}
RNDAV1=rtP_1+2 \cdot rtP_2+3 \cdot rtP_3+4 \cdot rtP_4=\sum_{t}
 rtAI(\hbox{tof})/N_{RND}
\end{equation}
is the average number of random ions detected in the presence of the true ions. Know we
know that
\begin{equation}\label{}
RNDAV2= \sum_{t} rtAI(\hbox{tof}) \cdot Palive(\hbox{tof})/N_{RND}
\end{equation}
is a much better estimate. Because of the dead time effects, $RNDAV2$ is smaller than
$RNDAV1$. So, it is better to use modified values of $rtP_{0,1,2,3,4}$.

Now we use a simple model to calculate the new values of $rtP_{0,1,2,3,4}$. $PDT$ is the
probability for the detector to be dead, because of the presence of the true ions. For
the sake of simplicity we assume that, $PDT$ does not depend on the number of ions
detected or on the TOF of an ion. The probability to be alive, is called $PAT=(1-PDT)$.
$PAT$ and $PDT$ can be calculated from $RNDAV1$ and $RNDAV2$:
\begin{eqnarray}
PAT &=& RNDAV2/RNDAV1 \nonumber \\
PDT &=& (1-PAT) \nonumber \\
\end{eqnarray}
$RP'_{0,1,2,3,4}$ are the modified values of $rtP_{0,1,2,3,4}$. They are related to the
measured values by:
\begin{eqnarray}
rtP_0' &=& rtP_0+rtP_1 \cdot PDT+rtP_2 \cdot PDT^2+rtP_3 \cdot PDT^3+rtP_4 \cdot PDT^4 \nonumber \\
rtP_1' &=& rtP_1 \cdot PAT+rtP_2 \cdot PDT \cdot PAT \cdot 2+rtP_3 \cdot PDT^2 \cdot PAT \cdot 3 \nonumber \\
~   &+& rtP_4 \cdot PDT^3 \cdot PAT \cdot 4 \nonumber \\
rtP_2' &=& rtP_2 \cdot PAT^2+rtP_3 \cdot PDT \cdot PAT^2 \cdot 3+rtP_4 \cdot PDT^2 \cdot PAT^2 \cdot 6 \nonumber \\
rtP_3' &=& rtP_3 \cdot PAT^3+rtP_4 \cdot PDT \cdot PAT^3 \cdot 4 \nonumber \\
rtP_4' &=& rtP_4 \cdot PAT^4 \nonumber \\
\end{eqnarray}
The whole data treatment described above stays the same. Only the modified values
$rtP_{0,1,2,3,4}'$ have to be used instead of $rtP_{0,1,2,3,4}$. Then also the dead time
effects of the ion detector will be considered.

\begin{table}
\caption{The symbols and their meaning}
\begin{tabular}{rl}
\hline
  $N_{e}$ & number of events with proper electron trigger \\
  $N_{RND}$ & number of events with proper random trigger \\
  etP$_{0,1,2,3,4}$ & probability to detect 0,1,2,3 or 4 ions after an electron trigger \\
  rtP$_{0,1,2,3,4}$ & probability to detect 0,1,2,3 or 4 ions after a random trigger \\
  TP$_{0,1,2,3,4}$ & probability to detect 0,1,2,3 or 4 true ions after an electron trigger \\
  P$_{0,1,2,3,4}$ & probability for 0,1,2,3 or 4 true ions to be present after an electron trigger \\
  PD & detection efficiency of ions \\
  AES(x) & spectrum of all electrons detected, Area(AES(x)) = $N_{e}$ \\
  ES$_{0,1,2,3,4}$(x) & spectra of electrons detected in coincidence with 0,1,2,3,4 ions\\
  ~ & e.g. Area(ES$_{2}$(x)) = $N_{e} \cdot $ etP$_{2}$ \\
  BES$_{1,2,3}$(x) & random contribution to electron spectra ES$_{0,1,2,3,4}$(x) \\
  TES$_{0,1,2,3}$(x) & true electron spectra, e.g. TES$_{2}$(x)=ES$_{2}$(x)-BES$_{2}$(x) \\
  Iregion(i) & TOF-interval that contains specific ions \\
  TES$_1$I(x,i) & true electron spectrum, coincident with ions in the interval Iregion(i) \\
  etAI(\hbox{tof}) & spectrum of all ions detected after electron triggers \\
  rtAI(\hbox{tof}) & spectrum of all ions detected after random triggers \\
  BetAI(\hbox{tof}) & random contribution to the ion spectrum etAI(\hbox{tof}) \\
  TetAI(\hbox{tof}) & true ion spectrum TetAI(\hbox{tof})=etAI(\hbox{tof})-BetAI(\hbox{tof}) \\
  etI(\hbox{tof}) & tof spectrum of the ions from single ion events triggered by electron detection \\
  etII(\hbox{tof1},\hbox{tof2}) & corresponding 2D ion TOF spectrum from ion pair events \\
  rtI(\hbox{tof}), RII(\hbox{tof1},\hbox{tof2}) & corresponding spectra for random triggers \\
  TetI(\hbox{tof}), BetI(\hbox{tof}) & true single ion spectrum and background \\
  TetII(\hbox{tof1},\hbox{tof2}), BetII(\hbox{tof1},\hbox{tof2}) & true ion pair 2D-spectrum and background \\
  IIregion(j) & region in the $\hbox{tof1}-\hbox{tof2}$ plane, that contains a specific ion pair \\
  CtsIIpair(j) & number of ion pairs detected inside IIregion(j) \\
  TCtsIIpair(j), BCtsIIpair(j) & number of true ion pairs inside IIregion(j) and and background  \\
  etEI(x,\hbox{tof}) & 2D electron position, ion tof spectrum from electron ion pair events \\
  TetEI(x,\hbox{tof}), BetEI(x,\hbox{tof}) & corresponding true spectrum and background \\
  ES$_2$IIpair(x,j) & electron spectrum, coincident with in ion pairs inside IIregion(j) \\
  TES$_2$IIpair(x,j), BES$_2$IIpair(x,j)& corresponding true electron spectrum and background \\
\end{tabular}
\label{symbol_table}
\end{table}

\clearpage




\begin{thebibliography}{00}


\bibitem{Lavolee99}
M. Lavol{'e}e, V. Brems, J. Chem. Phys. 110 (1999) 918.

\bibitem{Muramatsu02}
Y. Muramatsu, K. Ueda, N. Saito, H. Chiba, M. Lavoll{\'e}e, A.
Czasch, T. Weber, O. Jagutzki, H. Schmidt-B{\"o}cking, R. Moshammer,
U. Becker, K. Kubozuka, I. Koyano, Phys. Rev. Lett. 88 (2002)
133002.

\bibitem{Ueda05}
K. Ueda and J.H.D. Eland, J. Phys. B: At. Mol. Opt. Phys. 38 (2005)
S839; and references cited therein.


\bibitem{Heiser97}
F. Heiser, O. Gessner, J. Viefhaus, K. Wieliczek, R. Hentges, U.
Becker, Phys. Rev. Lett. 79 (1997) 2435.

\bibitem{Lafosse00}
A. Lafosse, M. Lebech, J. C. Brenot, P. M. Guyon, O. Jagutzki, L.
Spielberger, M. Vervloet, J. C. Houver, D. Dowek, Phys. Rev. Lett.
84 (2000) 5987.

\bibitem{Landers01}
A. Landers, Th. Weber, I. Ali, A. Cassimi, M. Hattaass, O. Jagutzki,
A. Nauert, T. Osipov, A. Staudte, M.H. Prior, H. Schmidt-B\"ocking,
R. D{\"o}rner, Phys. Rev. Lett. 87 (2001) 013002.


\bibitem{Defanis02}
A. De Fanis, N. Saito, A.A. Pavlychev, D.Yu. Ladonin, M. Machida, K.
Kubozuka, I. Koyano, K. Okada, K. Ikejiri, A. Cassimi, A. Czasch, R.
D{\"o}rner, H. Chiba, Y. Sato, K. Ueda, Phys. Rev. Lett. 89 (2002)
023006.

\bibitem{French}
C. Miron, M. Simon, N. Leclercq, D.L. Hansen, and P. Morin, Phys.
Rev. Lett. 81 (1998) 4104.

\bibitem{Kugeler04}
O. Kugeler, G. Pr\"umper, R. Hentges, J. Viefhaus, D. Rolles, U.
Becker, S. Marburger, U Hergenhahn, Phys. Rev. Lett. 93 (2004)
033002.

\bibitem{Pruemper05JPB}
G. Pr\"umper, Y. Tamenori, A. De Fanis, U. Hergenhahn, M. Kitajima,
M. Hoshino, H. Tanaka, K. Ueda, J. Phys. B: At. Mol. Opt. Phys. 38
(2005) 1.

\bibitem{Rolles05}
D. Rolles, M. Braune, S. Cvejanovic , O. Ge{\ss}ner, R. Hentges, S.
Korica, B. Langer, T. Lischke, G. Pruemper, A. Reinkoester, J.
Viefhaus, B. Zimmermann, V. McKoy, U. Becker, Nature 437 (2005) 711.

\bibitem{Kaemmerling92}
B. K\"ammerling, B Kr\"assigm V. Schmidt, J. Phys. B: At. Mol. Opt. Phys. 25 (2005)
3621.


\bibitem{Pruemper05JESRP}
G. Pr\"umper, K. Ueda, U. Hergenhahn, A. De Fanis, Y. Tamenori, M.
Kitajima, M. Hoshino, H. Tanaka, J. Electron Spectrosc. Relat.
Phenom. 144-147 (2005) 227.

\bibitem{Ohashi01NIMa}
H. Ohashi, E. Ishiguro, Y. Tamenori, H. Kishimoto, M. Tanaka, M.
Irie, T. Tanaka, T. Ishikawa, Nucl. Instrum. Methods Phys. Res. A
467-468 (2001) 529.

\bibitem{Ohashi01NIMb}
H. Ohashi, E. Ishiguro, Y. Tamenori, H. Okumura, A. Hiraya, H.
Yoshida, Y. Senda, K. Okada, N. Saito, I.H. Suzuki, K. Ueda, T.
Ibuki, S. Nagoka, I. Koyano, T. Ishikawa, Nucl. Instrum. Methods
Phys. Res. A 467-468 (2001) 533.

\bibitem{Tanaka96JSR}
T. Tanaka, H. Kitamura, J. Synchrotron Radiat. 3 (1996) 47.

\end{thebibliography}
\end{document}